\documentclass[conference]{IEEEtran}
\IEEEoverridecommandlockouts

\usepackage{cite}
\usepackage{amsmath,amssymb,amsfonts}
\usepackage{algorithmic}
\usepackage{graphicx}
\usepackage{textcomp}
\usepackage{xcolor}
\usepackage{subfigure}

\def\BibTeX{{\rm B\kern-.05em{\sc i\kern-.025em b}\kern-.08em
    T\kern-.1667em\lower.7ex\hbox{E}\kern-.125emX}}

\DeclareMathOperator*{\E}{\mathbb{E}}

\newtheorem{theorem}{Theorem}

\begin{document}
	\title{Parameter Estimation in a Noisy 1D Environment via Two Absorbing Receivers}
	\author{\IEEEauthorblockN{Xinyu Huang\IEEEauthorrefmark{1}, Yuting Fang\IEEEauthorrefmark{2}, Adam Noel\IEEEauthorrefmark{3}, and Nan Yang\IEEEauthorrefmark{1}}
		\IEEEauthorblockA{\IEEEauthorrefmark{1}Research School of Electrical, Energy and Materials Engineering, Australian National University, Canberra, ACT, Australia}
		\IEEEauthorblockA{\IEEEauthorrefmark{2}Department of Electrical and Electronic Engineering, University of Melbourne, Parkville, VIC, Australia}
		\IEEEauthorblockA{\IEEEauthorrefmark{3}School of Engineering, University of Warwick, Coventry, CV4 7AL, UK}
		\IEEEauthorblockA{Email: \{xinyu.huang1, nan.yang\}@anu.edu.au, yuting.fang@unimelb.edu.au, adam.noel@warwick.ac.uk}}

\maketitle

\begin{abstract}
This paper investigates the estimation of different parameters, e.g., propagation distance and flow velocity, by utilizing two fully-absorbing receivers (RXs) in a one-dimensional (1D) environment. The time-varying number of absorbed molecules at each RX and the number of absorbed molecules in a time interval as time approaches infinity are derived. Noisy molecules in this environment, that are released by sources in addition to the transmitter, are also considered. A novel estimation method, namely difference estimation (DE), is proposed to eliminate the effect of noise by using the difference of received signals at the two RXs. For DE, the Cramer-Rao lower bound (CRLB) on the variance of estimation is derived. Independent maximum likelihood estimation is also considered at each RX as a benchmark to show the performance advantage of DE. Aided by particle-based simulation, the derived analytical results are verified. Furthermore, numerical results show that DE attains the CRLB and is less sensitive to the change of noise than independent estimation at each RX.
\end{abstract}


\section{Introduction}

Nanoscale communication is critical for nanomachines to collaboratively execute complex tasks in biomedical applications, e.g., \textit{in vivo} drug delivery and surgery. Molecular communication (MC) is envisaged as one of the most promising methods for nano-scale communication, where molecules act as information carriers. MC possesses important characteristics, such as low energy consumption and potential for biocompatibility, which makes it more suitable for \textit{in vivo} applications than other nanoscale communication methods, e.g., electromagnetic methods \cite{farsad2016comprehensive}.

Parameter monitoring in the \textit{in vivo} environment is a potential application of MC, which is pivotal for healthcare, e.g., anomaly detection and targeted drug delivery \cite{chahibi2013molecular}. For example, estimating the emission rate of biomarkers by a tumor could help to diagnose the stage of tumor development. Moreover, distance estimation could help to accurately locate a tumor and deliver drugs to the site. Furthermore, estimating the flow velocity and degradation rate of emitted biomarkers could help to evaluate the blood velocity and pH near the tumor, respectively, since the degradation rate varies with pH as discussed in \cite{chang2005physical}. Motivated by these applications, some previous studies have provided valuable insights into parameter monitoring, e.g., \cite{noel2015joint,wang2015algorithmic,schafer2019eigenfunction,miao2019cooperative}. \cite{noel2015joint} and \cite{wang2015algorithmic} considered estimation by a single transparent receiver ($\mathrm{RX}$) and non-transparent $\mathrm{RX}$, respectively, where a transparent $\mathrm{RX}$ is one that does not interact with molecules. \cite{schafer2019eigenfunction} considered estimation in a one-dimensional (1D) pipe with two absorbing boundaries, where the estimation is based on particle concentration and flux in the pipe. \cite{miao2019cooperative} considered estimation by multiple non-transparent $\mathrm{RXs}$ based on data-fitting expressions of the channel impulse response (CIR), where the CIR is the expected number of absorbed molecules at each $\mathrm{RX}$ \cite{jamali2019channel}. Although these studies stand on their own merits, they did not analytically derive CIRs at multiple non-transparent $\mathrm{RXs}$ and apply such CIRs to estimate parameters in the MC environment.

In this paper, we, for the first time, perform parameter estimation by using analytically derived expressions of CIR at two dependent non-transparent $\mathrm{RXs}$. We consider a 1D environment, where a transmitter ($\mathrm{TX}$) continuously releases molecules into the environment. The 1D environment is worthy of investigation since it can approximate many practical biological channels such as capillaries, blood vessels, active transportation channels, and communication on bio-chips \cite{formaggia2003one,manocha2016dielectrophoretic}. Moreover, we emphasize that the 1D diffusion channel has been frequently applied to investigate several aspects of MC in some publications, e.g., \cite{schafer2019eigenfunction,varshney2018flow,chouhan2019optimal}. Furthermore, \cite{1907.04239} recently considered using approximative analytical expressions for hitting probabilities at two spherical fully-absorbing $\mathrm{RXs}$ in a three-dimensional (3D) environment.

The intended $\mathrm{TX}$ might not be the only source of molecules. We also consider other sources that secrete the same type of molecules as the intended $\mathrm{TX}$, which interfere with the parameter estimation at $\mathrm{RXs}$. In this paper, we refer to the molecules not emitted from the intended $\mathrm{TX}$ as noisy molecules. To reduce the impact of noisy molecules on the estimation, we propose a novel estimation method, namely, difference estimation (DE). In this method, the difference between the number of absorbed molecules at two fully-absorbing $\mathrm{RXs}$ is used to estimate parameters, where fully-absorbing $\mathrm{RXs}$ are those that absorb molecules once they hit the $\mathrm{RX}$ surface. In addition, we consider independent maximum-likelihood (ML) estimation at each $\mathrm{RX}$ as a benchmark to show performance superiority of DE.

Our major contributions are summarized as follows. We first derive the time-varying expression for the expected number of absorbed molecules at each $\mathrm{RX}$ when the $\mathrm{TX}$ emits molecules continuously. We next derive the asymptotic expression for the number of absorbed molecules within a time interval at each $\mathrm{RX}$ after sufficient time has passed. We further derive the Cramer-Rao lower bound (CRLB) on the variance of different parameters which include distances, emission rate, degradation rate of molecules, and flow velocity. Aided by particle-based simulation, we verify results and show that DE attains the CRLB. Importantly, our numerical results show that the impact of noise on the performance of DE is much less than that on independent estimation of each $\mathrm{RX}$, which demonstrates the superiority of DE in reducing the impact of noise.

% Our contributions are summarized as follows:
%\begin{enumerate}
%	\item We derive the expected number of absorbed bimarkers, i.e., the CIR, at each $\mathrm{ES}$ when the tumor emits biomarkers continuously based on our previous work in \cite{xinyu}. We further derive the asymptotic absorbed biomarkers within a time interval at each $\mathrm{ES}$ as time approaches infinity.
%	\item For independent estimation and DE, we derive the Cramer-Rao lower bound (CRLB) for the variance of estimated parameters.
%\end{enumerate}

%Aided by the particle-based simulation, we verify the accuracy of our analytical results. In addition, our numerical results show that both $\mathrm{ESs}$ attains the CRLB when the noise is zero, and DE attains the CRLB for any given noise. It also shows that the impact of noise on performance of DE is much less than that on independent estimation such that DE achieves a better performace than independent estimation at both $\mathrm{ESs}$.

\section{System Model}

\begin{figure}[!t]
\begin{center}
\includegraphics[width=1\columnwidth]{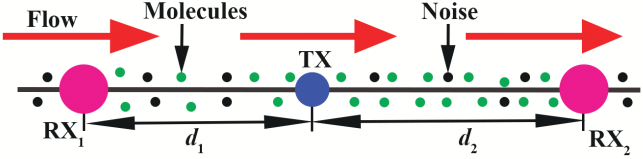}
\caption{Illustration of the system model, where one $\mathrm{TX}$ communicates with two fully-absorbing $\mathrm{RXs}$. A flow exists from $\mathrm{RX}_1$ towards $\mathrm{RX}_2$. Green dots denote molecules emitted from the $\mathrm{TX}$ while black dots denote noisy molecules.}\label{system_model_figure}\vspace{0em}
\end{center}\vspace{-2mm}
\end{figure}

In this paper, we consider a 1D unbounded environment, where a $\mathrm{TX}$ continuously releases molecules into the environment with a constant emission rate $\mu$ per second, as depicted in Fig. \ref{system_model_figure}. We assume that the the $\mathrm{TX}$ starts emission at $t=0\;\mathrm{s}$. To estimate environmental parameters, two fully-absorbing $\mathrm{RXs}$, i.e., $\mathrm{RX}_1$ and $\mathrm{RX}_2$, are placed at different sides of the $\mathrm{TX}$ with the distances $d_1$ and $d_2$ to the $\mathrm{TX}$, respectively. Moreover, we consider a steady uniform flow $\vec{v}$ in the direction from $\mathrm{RX}_1$ towards $\mathrm{RX}_2$, where the value of the uniform flow is $v\;(v>0)$. Once molecules are released from the $\mathrm{TX}$, they diffuse randomly with a constant diffusion coefficient $D$ and the constant flow. Additionally, molecules can degrade from the type $A$ into some other molecular species $\phi$, i.e., $A\stackrel{k}{\longrightarrow}\phi$ \cite[Ch. 9]{chang2005physical}, where $k$ is the degradation rate constant. The two $\mathrm{RXs}$ absorb type $A$ molecules as soon as they hit $\mathrm{RX}_1$ and $\mathrm{RX}_2$. In addition, we consider that the noisy molecules released by sources in addition to the intended $\mathrm{TX}$ may interfere with the estimation of environmental parameters. We emphasize that this additive noise is distinct from the randomness in the number of $A$ molecules absorbed by each $\mathrm{RX}$ due to diffusion, which may also interfere with the estimation. We note that one potential application of the system model in Fig. \ref{system_model_figure} is estimating parameters near a tumor in a blood-vessel environment.

In this system, estimation is based on the received signals of two $\mathrm{RXs}$ at time $t$, which is the number of absorbed molecules within the time interval $\left[t-\delta, t\right]$, where $\delta$ is the length of the time interval. We assume that each $\mathrm{RX}$ first performs $S$ observations separately, and then they perform estimation jointly based on the difference between observations at two $\mathrm{RXs}$. Therefore, we name this estimation method as DE. As discussed in \cite{fang2017convex}, the computation of difference can be conducted with the aid of another device. For example, the two $\mathrm{RXs}$ may transmit their observations to a fusion center that has easy access to computational resources. In this paper, we assume perfect transmission between each $\mathrm{RX}$ and the fusion center.

\begin{figure}[!t]
\begin{center}
\includegraphics[width=1\columnwidth]{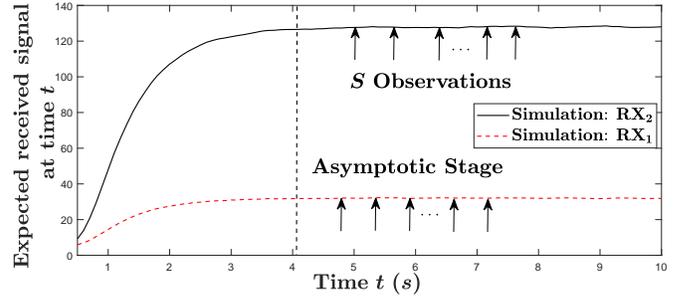}
\caption{The time-varying expected received signals, where $d_1=d_2=20\;\mu\mathrm{m}$, $v=6\;\mu\mathrm{m}/\mathrm{s}$, $D=79.4\;\mu\mathrm{m}^2/\mathrm{s}$, $k=0.8\;\mathrm{s}^{-1}$, $\mu=1000\;\mathrm{s}^{-1}$, and $\delta=0.5\;\mathrm{s}$. For other simulation details, please see Section \ref{NR}.}\label{segment}\vspace{0em}
\end{center}\vspace{-2mm}
\end{figure}

In Fig. \ref{segment}, we plot the expected received signals at two $\mathrm{RXs}$ in a noisy environment by the simulation. From this figure, we observe that the expected received signals become constant when $t>4\;\mathrm{s}$. We refer to the stage where the expected received signal is constant as the asymptotic stage. We note that it is a short time for two $\mathrm{RXs}$ to reach the asymptotic stage after the $\mathrm{TX}$ starts releasing molecules. Therefore, we assume that $\mathrm{RXs}$ start to make observations at the asymptotic stage as in \cite{fang2019expected}. One advantage for using asymptotic observations is that they do not depend on time, such that time synchronization is not required between devices, which improves the practicality of the proposed estimation method.

\section{Derivation of Channel Impulse Response}\label{DCIR}

In this section, we first derive the expected number of absorbed molecules at each $\mathrm{RX}$ until time $t$ due to continuous emission by the $\mathrm{TX}$. We then derive the closed-form expression for the number of absorbed molecules at each $\mathrm{RX}$ within the time interval $[t-\delta, t]$ as $t\rightarrow\infty$, which lays the foundation for parameter estimation in Section \ref{PE}.

We denote $N_j(t)$, $j=\{1,2\}$, as the expected number of absorbed molecules at $\mathrm{RX}_j$ by time $t$. Based on our previous work in \cite{xinyu}, we derive $N_j(t)$ in the following theorem:
\begin{theorem}\label{theore1}
The expected number of molecules absorbed at $\mathrm{RX}_j$ by time $t$ due to the continuous emission of molecules at the $\mathrm{TX}$ is given by
\begin{align}\label{N_1}
&N_j(t)=\mu\exp\left((-1)^j\frac{d_jv}{2D}\right)\sum_{i=0}^{\infty}\left[R_i(2(i+1)d+d_j,t,2)\right.\notag\\&\left.-R_i\left(2(i+2)d-d_j,t,2\right)-R_i\left(2id+d_j,t,0\right)\right.\notag\\&\left.+R_i(2(i+1)d-d_j,t,0)\right],
\end{align}
where $d=d_1+d_2$ is the distance between the two $\mathrm{RXs}$ and $R_i(x,t,a)$ is given by
\begin{align}\label{kaka}		&R_i(x,t,a)=\frac{\theta\kappa}{2}\left(\alpha\omega(t)-\hat{\alpha}\nu(t)\right)-\frac{i+1}{2}\left(\alpha\omega(t)+\hat{\alpha}\nu(t)\right)\notag\\&-\frac{\theta}{\sqrt{Dv^2+4kD}}\left(\hat{\alpha}\hat{\beta}(t)-\alpha\beta(t)\right)+(i+1)t.
\end{align}
In \eqref{kaka}, $\theta=d(i+1)(i+a)$, $\kappa=\sqrt{\frac{v^2}{4D^2}+\frac{k}{D}}$, $\alpha=\exp\left(x\kappa\right)$, $\hat{\alpha}=\exp\left(-x\kappa\right)$, $\beta(t)=\mathrm{erfc}\left(\frac{x}{\sqrt{4Dt}}+\sqrt{(k+\frac{v^2}{4D})t}\right)$, $\hat{\beta}(t)=\mathrm{erfc}\left(\frac{x}{\sqrt{4Dt}}-\sqrt{(k+\frac{v^2}{4D})t}\right)$, $\omega(t)=\int_{0}^{t}\beta(u)\mathrm{d}u$, and $\nu(t)=\int_{0}^{t}\hat{\beta}(u)\mathrm{d}u$. $\omega(t)$ and $\nu(t)$ are calculated numerically, e.g., using the built-in function \emph{integral} in MATLAB.
\end{theorem}
\begin{IEEEproof}
Please see Appendix \ref{app}.
\end{IEEEproof}

Since the received signal at each $\mathrm{RX}$ is the number of absorbed molecules within the time interval $[t-\delta, t]$, we express the received signal, denoted by $\Delta N_j(t)$, as
\begin{align}\label{Delta_t}
\Delta N_j(t)=N_j(t)-N_j(t-\delta).
\end{align}

As $t\rightarrow \infty$, we observe from Fig. \ref{segment} that $\Delta N_j(t)$ becomes constant. We derive the closed-form expression for the asymptotic value of $\Delta N_j(t)$, denoted by $\tilde{N}_j$, in the following theorem:
\begin{theorem}\label{theo2}
The asymptotic number of molecules absorbed by $\mathrm{RX}_j$ within an interval $[t-\delta, t]$ as $t\rightarrow\infty$, denoted by $\tilde{N}_j$, is derived as
\begin{align}\label{asy}
\tilde{N}_j=\mu\delta\exp\left((-1)^j\frac{d_jv}{2D}\right)
\frac{\exp\left(-d_j\kappa\right)-\exp\left((d_j-2d)\kappa\right)}{1-\exp\left(-2d\kappa\right)}.
\end{align}
\end{theorem}
\begin{IEEEproof}
Please see Appendix \ref{a2}.
\end{IEEEproof}

\section{Difference Estimation}\label{PE}

In this section, we assume that one environmental parameter is unknown and estimate this parameter by applying DE. For DE, we first derive the CRLB and then apply the method of moments to estimate the parameter. To show the performance advantage of DE, we consider independent ML estimation at each $\mathrm{RX}$ as a benchmark.

\subsection{Derivation of CRLB}

The CRLB is a lower bound on the variance of any unbiased estimator \cite[Ch. 3]{kay1993fundamentals}. For any unbiased estimator, the mean of estimated values equals its true value, which means that its mean squared error (MSE) equals the variance. An estimator is the minimum-variance unbiased (MVU) estimator if the MSE of the estimator attains the CRLB. Therefore, the CRLB can be applied to predict the performance of the MVU estimator.

To derive the CRLB, we need the joint conditional probability mass function (PMF) of $S$ observations of DE, where each observation is the difference between received signals at both $\mathrm{RXs}$. We first denote the vector that contains $S$ observations of received signal at $\mathrm{RX}_j$ by $\mathbf{g}_j=[g_{j,1}, g_{j,2},\ldots,g_{j,s},\ldots, g_{j,S}]$, where $g_{j,s}$ is the $s$th observation at $\mathrm{RX}_j$. We then denote the vector that contains the observations of DE by $\tilde{\mathbf{g}}=\left[\tilde{g}_1, \tilde{g}_2,\ldots, \tilde{g}_s,\ldots, \tilde{g}_S\right]$ where $\tilde{g}_s=g_{2,s}-g_{1,s}$. For the unknown parameter $\varepsilon$, we denote the joint conditional PMF of DE by $p\left(\tilde{\mathbf{g}}|\varepsilon\right)$ that are obtained by multiplication of the conditional PMF of each observation of DE if these observations are independent. To keep the independence of each observation in $\tilde{\mathbf{g}}$, we need to first guarantee the independence of observations in $\mathbf{g}_j$. In this paper, we assume that each observation in $\mathbf{g}_j$ is independent of other observations\footnote{Although we cannot guarantee perfect independence between successive observations in $\mathbf{g}_j$, the dependence can become extremely rare when the time between successive observations is sufficiently long.} and the time between two successive observations in $\mathbf{g}_j$ should be at least larger than $\delta$.

As we model the release time of each molecule at the $\mathrm{TX}$ as a continuous random process, the time interval between releasing two successive molecules is a random variable (RV). Thus, the release time of each molecule is different, which means that the received signal follows a Poisson binomial distribution since each molecule has a different probability of being absorbed by the time the observation is made. As the Poisson binomial distribution is cumbersome to work with, we approximate it by a Poisson distribution. The approximation becomes more accurate when the number of trials, i.e., emitted molecules, is larger and the success probability $\hat{P}_j(t)$ is smaller \cite{le1960approximation}. We assume that the number of absorbed noisy molecules within the time interval at each $\mathrm{RX}$ are identically and independently distributed Poisson RVs with constant mean $\xi$ as in \cite{noel2014optimal}. This assumption is reasonable since sources in addition to the intended $\mathrm{TX}$ can also be regarded as $\mathrm{TXs}$. Due to the additivity of two Poisson RVs, we model each observation of the received signal as a Poisson RV with mean $\tilde{N}_j+\xi$. As $g_{2,s}$ and $g_{1,s}$ follow a Poisson distribution, the difference between observations, i.e., $\tilde{g}_s$, follows a Skellam distribution with mean $\tilde{N}_2-\tilde{N}_1$. According to the PMF of Skellam distribution in \cite{karlis2003analysis}, the joint conditional PMF is
\begin{align}\label{p_g}
p\left(\tilde{\mathbf{g}}|\varepsilon\right)=&\prod_{s=1}^{S}
\exp\left(-\left(\hat{N}_1+\hat{N}_2\right)\right)
\left(\frac{\hat{N}_2}{\hat{N}_1}\right)^{\frac{\tilde{g}_s}{2}}\notag\\
&\times I_{\tilde{g}_s}\left(2\sqrt{\hat{N}_1\hat{N}_2}\right),
\end{align}
where $\hat{N}_j=\tilde{N}_j+\xi$ and $I_{\tilde{g}_s}(2\sqrt{\hat{N}_1\hat{N}_2})$ is the modified Bessel function of the first kind \cite{gradshteyn2014table}. According to \cite[Ch. 3]{kay1993fundamentals}, the CRLB of the variance of an unbiased estimator, denoted by $\mathrm{var}\left(\hat{\varepsilon}\right)$, is $\mathrm{var}(\hat{\varepsilon})\geq\frac{1}{L(\varepsilon)}$,
where $\hat{\varepsilon}$ is the estimated value of $\varepsilon$, $\mathrm{var}(\hat{\varepsilon})=\E\left[\left(\hat{\varepsilon}-\varepsilon\right)^2\right]$ for the unbiased estimator, $\E[\cdot]$ represents the expectation, and $L(\varepsilon)$ is the Fisher information that is given by \cite[eq. (3.6)]{kay1993fundamentals}
\begin{align}\label{I_i}
L(\varepsilon)=-\E\left[\frac{\partial^2\ln p(\tilde{\mathbf{g}}|\varepsilon)}{\partial\varepsilon^2}\right],
\end{align}
where $\E[\cdot]$ is taken with respect to $\tilde{g}_s$. We derive $L\left(\varepsilon\right)$ in the following theorem:
\begin{theorem}\label{thoe3}
The Fisher information of DE for unknown parameter $\varepsilon$ is derived as
\begin{align}\label{Lcn}
L(\varepsilon)=&S\Bigg[\frac{3\hat{N}_2-\hat{N}_1}{4\hat{N}_2^2}\gamma_2^2(\varepsilon)
+\frac{3\hat{N}_1-\hat{N}_2}{4\hat{N}_1^2}\gamma_1^2(\varepsilon)
-\frac{\hat{N}_1+\hat{N}_2}{2\hat{N}_1\hat{N}_2}\notag\\
&\times\gamma_1(\varepsilon)\gamma_2(\varepsilon)
+\left(\vartheta-\frac{4\hat{N}_2^2+3\hat{N}_2-\hat{N}_1}
{4\hat{N}_1\hat{N}_2}\right)\notag\\
&\times\Bigg(\sqrt{\frac{\hat{N}_2}{\hat{N}_1}}\gamma_1(\varepsilon)
+\sqrt{\frac{\hat{N}_1}{\hat{N}_2}}\gamma_2(\varepsilon)\Bigg)^2\Bigg],
\end{align}
where $\vartheta\!=\!\sum_{\tilde{g}_s=\zeta_1}^{\zeta_2}\!\!\frac{I^2_{\tilde{g}_s-1}
\left(2\sqrt{\hat{N}_1\hat{N}_2}\right)}{I_{\tilde{g}_s}
\left(2\sqrt{\hat{N}_1\hat{N}_2}\right)}\!\exp\!\left(-\!\left(\hat{N}_1
+\hat{N}_2\right)\!\right)\!\left(\frac{\hat{N}_2}
{\hat{N}_1}\right)\!^{\frac{\tilde{g}_s}{2}}$ and $\gamma_j(\varepsilon)=\frac{\partial\hat{N}_j}{\partial\varepsilon}$.
$\zeta_1$ and $\zeta_2$ are found numerically as explained in the Appendix \ref{a3}. In Table \ref{table}, we list $\gamma_j(\varepsilon)$ for the distance between the $\mathrm{TX}$ and $\mathrm{RX}_2$ $d_2$, flow velocity $v$, emission rate $\mu$ and degradation rate $k$.	
\end{theorem}
\begin{IEEEproof}
Please see Appendix \ref{a3}.
\end{IEEEproof}

\begin{table*}[!ht]
\newcommand{\tabincell}[2]{\begin{tabular}{@{}#1@{}}#2\end{tabular}}
\centering
\caption{$\gamma_j(\varepsilon)$ for each estimated parameter}\label{table}
\scalebox{0.9}{\begin{tabular}{|l|l|}\hline
$\gamma_j(d_2)$&\tabincell{c}{$\frac{\mu\Delta t\exp\left((-1)^j\frac{d_jv}{2D}\right)}{1-\exp\left(-2d\kappa\right)}\left[\exp\left(-d_j\kappa\right)\left(\frac{v}{2D}-(-1)^j\kappa\right)-\exp\left(\left(d_j-2d\right)\kappa\right)\left(\frac{v}{2D}+(-1)^j\kappa\right)\right]$}\\
\hline
\rule{0pt}{0.9pt}$\gamma_j(\mu)$&$\tilde{N}_j/\mu$\\
\hline
$\gamma_j(v)$&\tabincell{c}{$\frac{\mu\Delta t\exp\left((-1)^j\frac{d_jv}{2D}\right)}{\left(1-\exp\left(-2d\kappa\right)\right)^2}\left\{\left(\exp\left(2d\kappa\right)-1\right)\left[\exp\left(-(2d+d_j)\kappa\right)\left(\frac{(-1)^jd_j}{2D}-\frac{d_jv}{\sqrt{4v^2D^2+16D^3k}}\right)-\exp\left((d_j-4d)\kappa\right)\left(\frac{(-1)^jd_j}{2D}+\frac{(d_j-2d)v}{\sqrt{4D^2+v^2+16D^3k}}\right)\right]\right.$\\$\left.-\frac{vd}{\sqrt{v^2D^2+4kD^3}}\left[\exp\left(-(2d+d_j)\kappa\right)-\exp\left((d_j-2d)\kappa\right)\right]\right\}~~~~~~~~~~~~~~~~~~~~~~~~~~~~~~~~~~~~~~~~~~~~~~~~~~~~~~~~~~~~~~~~~~~~~~~~~~~~~~~~~~~~~~~~~~~~~~~~~~~~~~$}\\
\hline
$\gamma_j(k)$&\tabincell{c}{$\frac{\mu\Delta t\exp\left((-1)^j\frac{d_jv}{2D}\right)}{\left(1-\exp\left(-2d\kappa\right)\right)^2}\left[\exp\left(-(d_j+2d)\kappa\right)\left(\frac{d_j}{\sqrt{v^2+4Dk}}\left(1-\exp\left(2d\kappa\right)\right)-\frac{d}{\sqrt{\frac{v^2}{4D}+kD}}\right)-\exp\left((d_j-4d)\kappa\right)\left(\!\!\left(\exp\left(2d\kappa\right)-1\right)\frac{d_j-2d}{\sqrt{v^2+4kD}}-\frac{d}{\sqrt{\frac{v^2}{4}+kD}}\!\!\right)\!\!\right]$}\\
\hline
\end{tabular}}
\end{table*}

\subsection{Method of Moments}

For DE, we apply the method of moments to estimate the unknown parameter, which is finding $\hat{\varepsilon}$ that makes 1) the analytical expression for the mean of $\tilde{g}_s$, and 2) the mean of $\tilde{g}_s$ calculated by observed data, equal \cite[Ch. 9]{kay1993fundamentals}. As each item in the vector $\tilde{\mathbf{g}}$ is independent and follows the same distribution with the same mean, the mean of $\tilde{g}_s$ from observed data can be calculated by $\frac{1}{S}\sum_{s=1}^{S}\tilde{g}_s$. Since $\tilde{g}_s$ follows a Skellam distribution with mean $\tilde{N}_2-\tilde{N}_1$, we have the following estimation criteria:
\begin{align}\label{coess}
\left(\tilde{N}_2-\tilde{N}_1\right)\bigg|_{\varepsilon=\hat{\varepsilon}}=\frac{1}{S}\sum_{s=1}^{S}\tilde{g}_s,
\end{align}
where $\xi$ is not included such that the knowledge of noise from sources in addition to the intended $\mathrm{TX}$ is not required to perform the estimation.
\subsection{Benchmark}
To highlight the superiority of DE in eliminating the impact of noise, we consider independent estimation at each $\mathrm{RX}$ as a benchmark, where each $\mathrm{RX}$ performs ML estimation based on their $S$ observations in $\mathbf{g}_j$. We recall that each observation is approximated by a Poission distribution with mean $\hat{N}_j$. Therefore, the joint conditional PMF of $S$ observations at $\mathrm{RX}_j$, denoted by $p(\mathbf{g}_j|\varepsilon)$, is given by
$p\left(\mathbf{g}_j|\varepsilon\right)=\prod_{s=1}^{S}\hat{N}_j^{g_{j,s}}\exp\left(-\hat{N}_j\right)/g_{j,s}!$. The principle of ML estimation is finding $\hat{\varepsilon}_j$ that maximizes $p\left(\mathbf{g}_j|\varepsilon\right)$. By solving $\frac{\partial\ln p(\mathbf{g}_j|\varepsilon)}{\partial\varepsilon}=0$, we obtain the equivalent expression $\hat{N}_j\big|_{\varepsilon=\hat{\varepsilon}_j}=\frac{1}{S}\sum_{s=1}^{S}g_{j,s}$. Since $\xi$ is unknown in $\hat{N}_j$, one approach for independent estimation is ignoring the noise. By doing so, we obtain the following estimation criteria:
\begin{align}\label{ml}
\tilde{N}_j\big|_{\varepsilon=\hat{\varepsilon}_j}=\frac{1}{S}\sum_{s=1}^{S}g_{j,s}.
\end{align}

We highlight that \eqref{ml} indicates that the ML estimation at $\mathrm{RX}_j$ is to find $\hat{\varepsilon}_j$ that makes $\tilde{N}_j\big|_{\varepsilon=\hat{\varepsilon}_j}$ and the mean of $S$ observations in $\mathbf{g}_j$ equal.

\section{Numerical Results}\label{NR}
In this section, we present numerical results to validate our theoretical analysis and provide insightful discussions. Particle-based simulation is used to simulate the random propagation of molecules \cite{arifler2017monte}. The simulation time step is $t_\mathrm{sim}=0.001\;\mathrm{s}$ and all results are averaged over 2000 realizations. Throughout this section, we set $d=40\;\mu\mathrm{m}$, $D=79.4\;\mu\mathrm{m^2/s}$, $\mu=1000\;\mathrm{s^{-1}}$, $k=0.8\;\mathrm{s^{-1}}$, and $\delta=0.5\;\mathrm{s}$ \cite{deng2017analyzing}, unless otherwise stated.

\begin{figure}[!t]
\begin{center}	\includegraphics[height=2in,width=0.8\columnwidth]{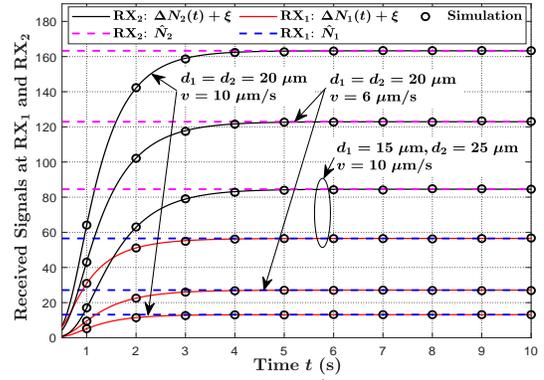}\vspace{-1em}
\caption{Received signals $\Delta N_j(t)+\xi$ and $\hat{N}_j$ versus time $t$ at $\mathrm{RX}_j$ for three parameter sets: 1) $d_1=d_2=20\;\mu\mathrm{m}$, $v=10\;\mu\mathrm{m/s}$, 2) $d_1=d_2=20\;\mu\mathrm{m}$, $v=6\;\mu\mathrm{m/s}$, 3) $d_1=15\;\mu\mathrm{m}, d_2=25\;\mu\mathrm{m}$, $v=6\;\mu\mathrm{m/s}$, where $\xi=5$ for all parameter sets.}\label{received_sig}\vspace{0em}
\end{center}
\end{figure}

In Fig. \ref{received_sig}, we plot the received signals $\Delta N_j(t)+\xi$ and the asymptotic received signals $\hat{N}_j$ at two $\mathrm{RXs}$ versus time $t$, where different flow velocities and distances between the $\mathrm{TX}$ and two $\mathrm{RXs}$ are considered to investigate the impact of the flow and distance on the received signals\footnote{The P$\mathrm{\acute{e}}$clet number is defined as the ratio between the advection and diffusion, which is $\mathrm{Pe}=\frac{vd_2}{D}$. In each parameter set, $\mathrm{Pe}>1$. Therefore, flow is the dominant transport mechanism in this environment \cite{jamali2019channel}.}. We first observe that the simulation matches well with $\Delta N_j(t)+\xi$ and $\hat{N}_j$, which demonstrates the correctness of \eqref{N_1} and \eqref{asy}. Furthermore, comparing parameter sets 1) and 2), we observe that the received signals increase at $\mathrm{RX}_2$ and decrease at $\mathrm{RX}_1$ with an increase in the flow velocity. This is because we assume that the direction of the flow is from $\mathrm{RX}_1$ towards $\mathrm{RX}_2$. In addition, comparing parameter sets 2) and 3), we observe that the received signals increase with a decrease in the distance between the $\mathrm{TX}$ and $\mathrm{RX}$.
\begin{figure}[!t]
\begin{center}
\includegraphics[height=4.2in,width=1\columnwidth]{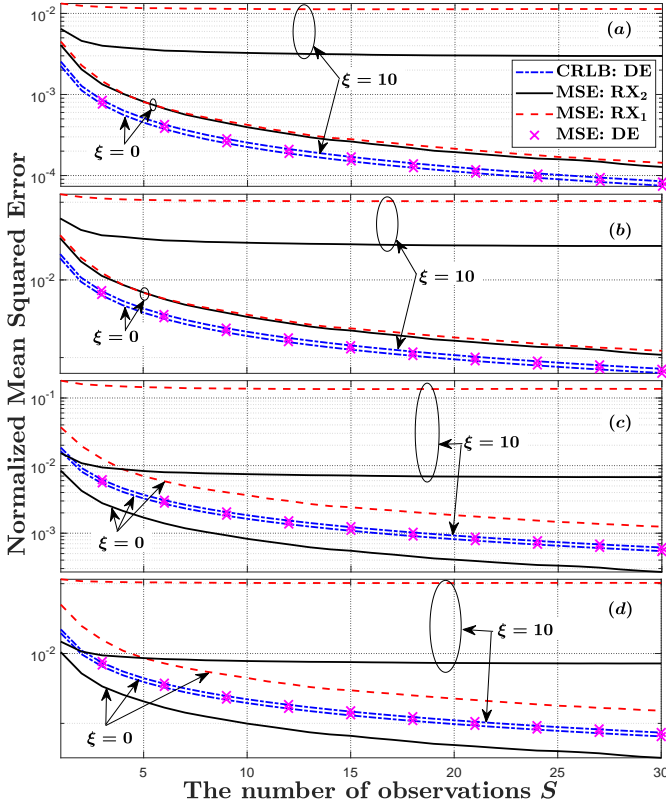}\vspace{-1em}
\caption{The normalized MSE versus the number of observations $S$, where $\xi=0$ and $\xi=10$ are considered. $(a)$: Estimation of $d_2$. $(b):$ Estimation of $v$. $(c):$ Estimation of $\mu$. $(d):$ Estimation of $k$.}\label{d2v}\vspace{0em}
\end{center}
\end{figure}

In Fig. \ref{d2v}, we plot the normalized MSE for DE and two $\mathrm{RXs}$ and the CRLB for DE versus the number of observations, where $\xi=0$ and $\xi=10$ in this environment are considered. In this figure, we set $d_1=d_2=20\;\mu\mathrm{m}$ and $v=6\;\mu\mathrm{m}/\mathrm{s}$. Both MSE and CRLB are normalized over $\varepsilon^2$. First, we observe that the normalized MSE attains the CRLB for DE, which indicates that the method of moments applied in DE achieves the minimum variance. Second, we observe that the normalized MSEs at two $\mathrm{RXs}$ increase with the increase in noise. %The impact of the increasing noise is higher on $\mathrm{RX}_1$ than $\mathrm{RX}_2$ since the number of absorbed molecules at $\mathrm{RX}_1$ is much smaller than that at $\mathrm{RX}_2$.
Third, in Fig. $4(a)$ and Fig. $4(b)$, we observe that the normalized MSEs at two $\mathrm{RXs}$ are close to each other when $\xi=0$ and the MSE of DE is always lower than that at either of the $\mathrm{RXs}$. This is because $d_2$ and $v$ have no impact on the number of molecules in this environment, which means that the performance of the estimation for $d_2$ and $v$ is not related to the number of molecules absorbed at each $\mathrm{RX}$. As DE combines observations of both $\mathrm{RXs}$, DE achieves a better performance. When noise increases, the impact on the estimation of DE is much less than that at either $\mathrm{RX}$, which indicates the superiority of DE in eliminating the impact of noise in estimation. This is because the difference of received signals in DE offsets the impact of noise. Fourth, in Fig. $4(c)$ and Fig. $4(d)$, we observe a large gap between normalized MSEs of either $\mathrm{RX}$ and the performance of $\mathrm{RX}_2$ is better than DE when $\xi=0$. This is because $\mu$ and $k$ influence the number of molecules in the environment, which means that the number of absorbed molecules at either $\mathrm{RX}$ influences the performance of estimation. When $\xi=0$, $\mathrm{RX}_2$ performs estimation based on a larger number of molecules than DE and $\mathrm{RX}_1$. Therefore, $\mathrm{RX}_2$ achieves the best performance. It is worth noting that the performance of $\mathrm{RX}_2$ becomes worse than DE for $S>2$ when $\xi=10$ and the increase of MSE in DE is much less than the increase at $\mathrm{RX}_2$, which still proves that DE is superior than independent estimation in eliminating the impact of noise. This superiority becomes obvious with the increase in $\xi$ and $S$.

\begin{figure}[!t]
\begin{center}
\includegraphics[height=1.8in,width=1\columnwidth]{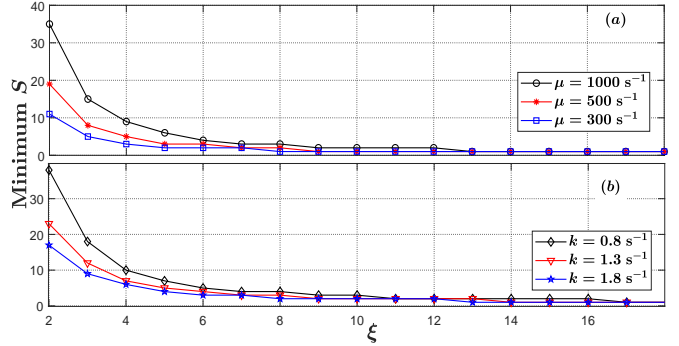}\vspace{-1em}
\caption{The minimum number of observations $S$ required for MSE in DE to be less than the MSE at $\mathrm{RX}_2$ versus the mean of noise $\xi$. $(a)$: Estimation of $\mu$. $(b)$: Estimation of $k$.}\label{observations}\vspace{-0em}
\end{center}
\end{figure}

In Fig. \ref{observations}, we plot the minimum number of observations required for the MSE in DE to be less than the MSE at $\mathrm{RX}_2$, where the estimation of $\mu$ and $k$ are considered because of the observations of Fig. $4(c)$ and Fig. $4(d)$. From this figure, we first observe that the minimum number of observations required decreases with an increase in the noise mean, which indicates that the increase of MSE at DE is much slower than that at $\mathrm{RX}_2$, demonstrating that DE is less sensitive to the increase in noise than $\mathrm{RX}_2$. After $\xi>13$ and $\xi>17$ for the estimation of $\mu$ and $k$, respectively, the MSE in DE is always less than the MSE at $\mathrm{RX}_2$ since the minimum $S$ drops to one. Second, we observe that the minimum number of observations decreases with a decrease in $\mu$ or increase in $k$. This is because a decrease in $\mu$ or increase in $k$ results in a decrease in the received signal at $\mathrm{RX}_2$, which makes $\mathrm{RX}_2$ more susceptible to an increase in noise than DE.

\section{Conclusion}
In this paper, we investigated parameter estimation by the cooperation of two $\mathrm{RXs}$ in a noisy 1D environment. We derived the analytical expressions for the CIR at both $\mathrm{RXs}$. Moreover, we considered DE and derived the CRLB. To show the advantage of DE, independent estimation at each $\mathrm{RX}$ was also investigated. Our numerical results verified our analytical results. Numerical results also showed that DE is less sensitive to an increase in noise than independent estimation. For the estimation of $d_2$ and $v$, DE always achieves better performance than independent estimation. For the estimation of $\mu$ and $k$, DE becomes better than independent estimation with an increase in noise or in the number of observations. Future work includes considering mobile $\mathrm{RXs}$ and extending the estimation to a 3D environment.

\appendices

\section{Proof of Theorem \ref{theore1}}\label{app}

We express $N_j(t)$ as $N_j(t)=\mu\int_{0}^{t}\hat{P}_j(u)\mathrm{d}u$, where $\hat{P}_j(t)$ is the fraction of absorbed molecules by time $t$ for an impulsive emission at the $\mathrm{TX}$. To obtain $N_j(t)$, we first calculate $\hat{P}_j(t)$.

We denote $f_\mathrm{v}(d,t,k,v)$ and $f(d,t,k)$ as the hitting rate with and without flow, respectively, when only one $\mathrm{RX}$ exists. According to \cite{s2012molecular}, $f_\mathrm{v}(d,t,k,v)$ is given by
\begin{align}\label{hat_f}
f_\mathrm{v}(d,t,k,v)=\frac{d}{\sqrt{4\pi Dt^3}}\exp\!\left(\frac{dv}{2D}\!-\!\frac{d^2}{4Dt}\!-\!\left(\frac{v^2}{4D}
\!+\!k\right)t\right),
\end{align}
where $d$ is the distance between the $\mathrm{TX}$ and $\mathrm{RX}$. In \eqref{hat_f}, we assume that the direction of the flow is from the $\mathrm{TX}$ toward $\mathrm{RX}$. For the opposite direction, we replace $v$ with $-v$.

By setting $v=0$ in \eqref{hat_f}, we obtain $f(d,t,k)$ as
\begin{align}\label{f}
	f(d,t,k)=\frac{d}{\sqrt{4\pi Dt^3}}\exp\left(-\frac{d^2}{4Dt}-kt\right).
\end{align}

Based on \eqref{hat_f} and \eqref{f}, we obtain
\begin{align}\label{f_v2}
	f_\mathrm{v}(d,t,k,v)=\exp\left(\frac{dv}{2D}\right)f\left(d,t,k+\frac{v^2}{4D}\right).
\end{align}

We denote $F_\mathrm{v}(d,t,k,v)$ and $F(d,t,k)$ as the fraction of absorbed molecules by time $t$ with and without flow, respectively, when only one $\mathrm{RX}$ exists. which are obtained by integrating $f_\mathrm{v}(d,t,k,v)$ and $f(d,t,k)$ over time $t$. Therefore, they have the same relationship as \eqref{f_v2}, which is
\begin{align}\label{F_v2}
F_\mathrm{v}(d,t,k,v)=\exp\left(\frac{dv}{2D}\right)F\left(d,t,k+\frac{v^2}{4D}\right).
\end{align}

Similar to the method in \cite[Sec. III]{xinyu}, we can express $\hat{P}_1(t)$ and $\hat{P}_2(t)$ as \cite[eqs. (12), (13)]{xinyu}
\begin{align}\label{p2}
	\hat{P}_2(t)=F_\mathrm{v}(d_2,t,k,v)-\hat{P}_1(t)*f_\mathrm{v}(d_1+d_2,t,k,v),
\end{align}
\begin{align}\label{p1}
	\hat{P}_1(t)=F_\mathrm{v}(d_1,t,k,-v)-\hat{P}_2(t)*f_\mathrm{v}(d_1+d_2,t,k,-v),
\end{align}
where $*$ stands for convolution. Substituting \eqref{f_v2} and \eqref{F_v2} into \eqref{p1} and \eqref{p2} and performing the Laplace transform for \eqref{p1} and \eqref{p2}, we obtain
\begin{align}\label{lap_p}
     \hat{\mathcal{P}}_2(s)=\exp\left(\frac{d_2v}{2D}\right)\mathcal{P}_2\left(s,\frac{v^2}{4D}+k\right),
\end{align}
where $\hat{\mathcal{P}}_2(s)$ is the Laplace transform of $\hat{P}_2(t)$ and $\mathcal{P}_2(s,k)$ is the Laplace transform of $P_2(t,k)$ that is given by \cite[eq. (8)]{xinyu}. Performing the inverse Laplace transform of \eqref{lap_p}, we obtain $\hat{P}_2(t)=\exp\left(\frac{d_2v}{2D}\right)P_2(t,\frac{v^2}{4D}+k)$. Substituting $\hat{P}_2(t)$ into $N_2(t)=\mu\int_{0}^{t}\hat{P}_2(u)\mathrm{d}u$, we obtain $N_2(t)$. $N_1(t)$ can be obtained by exchanging $d_1$ and $d_2$ and replacing $v$ with $-v$ therein for $N_2(t)$. Based on the expressions for $N_1(t)$ and $N_2(t)$, a unified formula can be written as \eqref{N_1}.

\section{Proof of Theorem \ref{theo2}}\label{a2}

According to the final value theorem, if $\hat{P}_2(t)$ has a finite limit as $t\rightarrow\infty$, we have
\begin{align}\label{fvt}
\lim\limits_{t\rightarrow\infty}\hat{P}_2(t)=\lim\limits_{s\rightarrow 0}s\hat{\mathcal{P}}(s).
\end{align}

Substituting \eqref{lap_p} into \eqref{fvt}, we obtain $\hat{P}_2\left(t\right)\big|_{t\rightarrow\infty}$ as
\begin{align}\label{P_inf}
\hat{P}_2(t)\big|_{t\rightarrow\infty}&=\exp\left(\frac{d_2v}{2D}\right)\lim\limits_{s\rightarrow 0}s\mathcal{P}_2(s,\frac{v^2}{4D}+k)\notag\\&=\exp\left(\frac{d_2v}{2D}\right)P_{2,\mathrm{asy}}\left(k+\frac{v^2}{4D}\right),
\end{align}
where $P_{2,\mathrm{asy}}(k)$ is given by \cite[eq. (11)]{xinyu}\footnote{Please note that $k=0\;\mathrm{s}^{-1}$ is not considered in this paper.}. According to $N_2(t)=\mu\int_{0}^{t}\hat{P}_2(u)\mathrm{d}u$, absorbed molecules at $\mathrm{RX}_2$ within the time interval $[t-\delta, t]$ at the asymptotic stage is
\begin{align}\label{N_asy}
\tilde{N}_2=\mu\delta\hat{P}_2(t)\big|_{t\rightarrow\infty}.
\end{align}

Substituting \eqref{P_inf} into \eqref{N_asy}, we obtain $\tilde{N}_2$. $\tilde{N}_1$ can be obtained by exchanging $d_1$ and $d_2$ and replacing $v$ with $-v$ therein for $\tilde{N}_2$. Based on the expressions for $\tilde{N}_1$ and $\tilde{N}_2$, a unified formula can be written as \eqref{asy}.

\section{Proof of Theorem \ref{thoe3}}\label{a3}

For the CRLB to exist, the regularity condition \cite{kay1993fundamentals} must be satisfied, which is $\E\left[\frac{\partial\ln p\left(\tilde{\mathbf{g}}|\varepsilon\right)}{\partial\varepsilon}\right]=0$. Substituting \eqref{p_g} into $\E\left[\frac{\partial\ln p\left(\tilde{\mathbf{g}}|\varepsilon\right)}{\partial\varepsilon}\right]$, we obtain
\begin{align}\label{EE}
&\hspace{0mm}\E\left[\frac{\partial\ln p\left(\tilde{\mathbf{g}}|\varepsilon\right)}{\partial\varepsilon}\right]\!
=\!\sum_{s=1}^{S}\!-\!\gamma_1(\varepsilon)\!-\!\gamma_2(\varepsilon)\!
+\!\frac{\E\left[\tilde{g}_s\right]}{2}\!\left(\!\frac{\gamma_2(\varepsilon)}
{\hat{N}_2}\!-\!\frac{\gamma_1(\varepsilon)}{\hat{N}_1}\!\right)\notag\\
&+\left(\E\left[\frac{I_{\tilde{g}_s-1}\left(2\sqrt{\hat{N}_1\hat{N}_2}\right)}
{I_{\tilde{g}_s}\left(2\sqrt{\hat{N}_1\hat{N}_2}\right)}\right]
-\frac{\tilde{g}_s}{2\sqrt{\hat{N}_1\hat{N}_2}}\right)\notag\\
&\times\left(\sqrt{\frac{\hat{N}_2}{\hat{N}_1}}\gamma_1(\varepsilon)
+\sqrt{\frac{\hat{N}_1}{\hat{N}_2}}\gamma_2(\varepsilon)\right).
\end{align}

Based on the definition of expectation, $\E\left[\frac{I_{\tilde{g}_s-1}\left(2\sqrt{\hat{N}_1\hat{N}_2}\right)}{I_{\tilde{g}_s}\left(2\sqrt{\hat{N}_1\hat{N}_2}\right)}\right]$ is calculated as
%\begin{align}\label{exp}
%&\E\left[\frac{I_{\tilde{g}_s-1}\left(2\sqrt{\hat{N}_1\hat{N}_2}\right)}{I_{\tilde{g}_s}\left(2\sqrt{\hat{N}_1\hat{N}_2}\right)}\right]=\sum_{\tilde{g}_s=-\infty}^{\infty}\frac{I_{\tilde{g}_s-1}\left(2\sqrt{\hat{N}_1\hat{N}_2}\right)}{I_{\tilde{g}_s}\left(2\sqrt{\hat{N}_1\hat{N}_2}\right)}\notag\\&\times\exp\left(-\left(\hat{N}_1+\hat{N}_2\right)\right)\left(\frac{\hat{N}_2}{\hat{N}_1}\right)^{\frac{\tilde{g}_s}{2}}I_{\tilde{g}_s}\left(2\sqrt{\hat{N}_1\hat{N}_2}\right)\notag\\&=\sqrt{\frac{\hat{N}_2}{\hat{N}_1}}\underbrace{\sum_{\tilde{g}_s=-\infty}^{\infty}\exp\left(-\left(\hat{N}_1+\hat{N}_2\right)\right)\left(\frac{\hat{N}_2}{\hat{N}_1}\right)^{\frac{\tilde{g}_s-1}{2}}I_{\tilde{g}_s-1}\left(2\sqrt{\hat{N}_1\hat{N}_2}\right)}_{\eta}\notag\\&=\sqrt{\frac{\hat{N}_2}{\hat{N}_1}},
%\end{align}
\begin{align}\label{exp}
&\E\left[\frac{I_{\tilde{g}_s-1}\left(2\sqrt{\hat{N}_1\hat{N}_2}\right)}
{I_{\tilde{g}_s}\left(2\sqrt{\hat{N}_1\hat{N}_2}\right)}\right]=
\sum_{\tilde{g}_s=-\infty}^{\infty}\frac{I_{\tilde{g}_s-1}
\left(2\sqrt{\hat{N}_1\hat{N}_2}\right)}{I_{\tilde{g}_s}
\left(2\sqrt{\hat{N}_1\hat{N}_2}\right)}\notag\\
&\times\exp\left(-\left(\hat{N}_1+\hat{N}_2\right)\right)
\left(\frac{\hat{N}_2}{\hat{N}_1}\right)^{\frac{\tilde{g}_s}{2}}
I_{\tilde{g}_s}\left(2\sqrt{\hat{N}_1\hat{N}_2}\right)\notag\\
&=\sqrt{\frac{\hat{N}_2}{\hat{N}_1}}\eta=\sqrt{\frac{\hat{N}_2}{\hat{N}_1}},
\end{align}
where
\begin{align}
\eta=&\sum_{\tilde{g}_s=-\infty}^{\infty}
\exp\left(-\left(\hat{N}_1+\hat{N}_2\right)\right)
\left(\frac{\hat{N}_2}{\hat{N}_1}\right)^{\frac{\tilde{g}_s-1}{2}}\notag\\
&\times I_{\tilde{g}_s-1}\left(2\sqrt{\hat{N}_1\hat{N}_2}\right)
\end{align}
is the summation of PMF of Skellam distribution. Therefore, $\eta=1$. Substituting \eqref{exp} and $\E\left[\tilde{g}_s\right]=\tilde{N}_2-\tilde{N}_1$ into \eqref{EE}, we obtain $\E\left[\frac{\partial\ln p\left(\tilde{\mathbf{g}}|\varepsilon\right)}{\partial\varepsilon}\right]=0$.

We then calculate $L(\varepsilon)$. Substituting \eqref{p_g} into \eqref{I_i}, we obtain
\begin{align}\label{Lc}
&L(\varepsilon)=\sum_{\tilde{g}_s=1}^{S}
-\frac{3\hat{N}_2-\hat{N}_1}{4\hat{N}_2^2}\gamma_2^2(\varepsilon)
+\frac{3\hat{N}_1-\hat{N}_2}{4\hat{N}_1^2}\gamma_1^2(\varepsilon)\notag\\
&-\frac{\hat{N}_1+\hat{N}_2}{2\hat{N}_1\hat{N}_2}\gamma_1(\varepsilon)
\gamma_2(\varepsilon)\left(\frac{\E\!\left[\frac{\tilde{g}_s
I_{\tilde{g}_s-1}}{I_{\tilde{g}_s}}\right]}
{\hat{N}_1\hat{N}_2}-\frac{\E\!\left[\tilde{g}_s^2\right]}{4\hat{N}_1\hat{N}_2}
+\frac{1}{2}\right.\notag\\
&\left.+\frac{1}{4}\!\left(\!\E\!\left[\!\frac{I_{\tilde{g}_s-2}
\left(\sqrt{2\hat{N}_1\hat{N}_2}\right)}{I_{\tilde{g}_s}
\left(\sqrt{2\hat{N}_1\hat{N}_2}\right)}\!\right]
+\E\!\left[\!\frac{I_{\tilde{g}_s+2}\left(\sqrt{2\hat{N}_1\hat{N}_2}\right)}
{I_{\tilde{g}_s}\left(\sqrt{2\hat{N}_1\hat{N}_2}\right)}\!\right]\!\right)\right.\notag\\
&\left.-\E\!\left[\!\frac{I^2_{\tilde{g}_s-1}\left(\sqrt{2\hat{N}_1\hat{N}_2}\right)}
{I^2_{\tilde{g}_s}\left(\sqrt{2\hat{N}_1\hat{N}_2}\right)}\!\right]\right)
\!\left(\!\sqrt{\frac{\hat{N}_2}{\hat{N}_1}}\gamma_1(\varepsilon)
+\sqrt{\frac{\hat{N}_1}{\hat{N}_2}}\gamma_2(\varepsilon)\!\right)^2.
\end{align}

Similar to the method in \eqref{exp}, we calculate $\E\left[\frac{\tilde{g}_sI_{\tilde{g}_s-1}}{I_{\tilde{g}_s}}\right]=\sqrt{\frac{\hat{N}_2}{\hat{N}_1}}\left(\hat{N}_2-\hat{N}_1+1\right)$, $\E\left[\tilde{g}_s^2\right]=\left(\hat{N}_2-\hat{N}_1\right)^2+\hat{N}_1+\hat{N}_2$, $\E\left[\frac{I_{\tilde{g}_s-2}\left(2\sqrt{\hat{N}_1\hat{N}_2}\right)}{I_{\tilde{g}_s}\left(2\sqrt{\hat{N}_1\hat{N}_2}\right)}\right]=\frac{\hat{N}_2}{\hat{N}_1}$, and $\E\left[\frac{I_{\tilde{g}_s+2}\left(2\sqrt{\hat{N}_1\hat{N}_2}\right)}{I_{\tilde{g}_s}\left(2\sqrt{\hat{N}_1\hat{N}_2}\right)}\right]=\frac{\hat{N}_1}{\hat{N}_2}$. In $L(\varepsilon)$, $\vartheta$ represents the value of $E\left[\frac{I^2_{\tilde{g}_s-1}\left(\sqrt{2\hat{N}_1\hat{N}_2}\right)}{I^2_{\tilde{g}_s}\left(\sqrt{2\hat{N}_1\hat{N}_2}\right)}\right]$ that is also calculated based on the definition of the expectation. For calculating $\vartheta$, we do not need the summation from $\tilde{g}_s=-\infty$ to $\tilde{g}_s=\infty$, and we focus on the range of $\tilde{g}_s$ that makes the PMF of the Skellam distribution larger than a threshold that we choose to be $10^{-3}$ in our paper, i.e., $\exp\left(-\left(\hat{N}_1+\hat{N}_2\right)\right)\left(\frac{\hat{N}_2}{\hat{N}_1}\right)^{\frac{\tilde{g}_s}{2}}I_{\tilde{g}_s}\left(2\sqrt{\hat{N}_1\hat{N}_2}\right)\geq10^{-3}$. The validation of this threshold is performed in our numerical tests. After solving this expression, we have $\zeta_1\leq\tilde{g}_s\leq\zeta_2$. Substituting $\E\left[\frac{\tilde{g}_sI_{\tilde{g}_s-1}}{I_{\tilde{g}_s}}\right]$, $\E\left[\tilde{g}_s^2\right]$, $\E\left[\frac{I_{\tilde{g}_s-2}\left(2\sqrt{\hat{N}_1\hat{N}_2}\right)}{I_{\tilde{g}_s}\left(2\sqrt{\hat{N}_1\hat{N}_2}\right)}\right]$, $\E\left[\frac{I_{\tilde{g}_s+2}\left(2\sqrt{\hat{N}_1\hat{N}_2}\right)}{I_{\tilde{g}_s}\left(2\sqrt{\hat{N}_1\hat{N}_2}\right)}\right]$, and $\vartheta$ into \eqref{Lc}, we obtain \eqref{Lcn}.

%Similar to the method in \eqref{exp}, we have $\E\left[\frac{\tilde{g}_pI_{\tilde{g}_p-1}}{I_{\tilde{g}_p}}\right]=\sqrt{\frac{\hat{N}_2}{\hat{N}_1}}\left(\hat{N}_2-\hat{N}_1+1\right)$, $\E\left[\tilde{g}_p^2\right]=\left(\hat{N}_2-\hat{N}_1\right)^2+\hat{N}_1+\hat{N}_2$, $\E\left[\frac{I_{\tilde{g}_p-2}\left(\sqrt{2\hat{N}_1\hat{N}_2}\right)}{I_{\tilde{g}_p}\left(\sqrt{2\hat{N}_1\hat{N}_2}\right)}\right]=\frac{\hat{N}_2}{\hat{N}_1}$, and $\E\left[\frac{I_{\tilde{g}_p+2}\left(\sqrt{2\hat{N}_1\hat{N}_2}\right)}{I_{\tilde{g}_p}\left(\sqrt{2\hat{N}_1\hat{N}_2}\right)}\right]=\frac{\hat{N}_1}{\hat{N}_2}$. We note that $E\left[\frac{I^2_{\tilde{g}_p-1}\left(\sqrt{2\hat{N}_1\hat{N}_2}\right)}{I^2_{\tilde{g}_p}\left(\sqrt{2\hat{N}_1\hat{N}_2}\right)}\right]$ is also calculated based on the definition of the expectation, which is expressed as $\vartheta$. For calculating $\vartheta$, we do not need the summation from $\tilde{g}_p=-\infty$ to $\tilde{g}_p=\infty$, and we just focus on the range of $\tilde{g}_p$ that makes the PMF of Skellam distribution larger than a threshold that is $10^{-3}$ in our paper. The validation of this threshold is performed from our numerical tests. Substituting $\E\left[\frac{\tilde{g}_pI_{\tilde{g}_p-1}}{I_{\tilde{g}_p}}\right]$, $\E\left[\tilde{g}_p^2\right]$, $\E\left[\frac{I_{\tilde{g}_p-2}\left(\sqrt{2\tilde{N}_1\tilde{N}_2}\right)}{I_{\tilde{g}_p}\left(\sqrt{2\tilde{N}_1\tilde{N}_2}\right)}\right]$, $\E\left[\frac{I_{\tilde{g}_p+2}\left(\sqrt{2\tilde{N}_1\tilde{N}_2}\right)}{I_{\tilde{g}_p}\left(\sqrt{2\tilde{N}_1\tilde{N}_2}\right)}\right]$
\bibliographystyle{IEEEtran}
\bibliography{refs}
\end{document}